\documentclass[prb,reprint,twocolumn]{revtex4}

\usepackage{graphicx}
\usepackage{dcolumn}
\usepackage{color}
\usepackage{amsmath,amsthm,amssymb,bm}
\usepackage{comment}

\begin{document}

\title{Control of Quantum Dynamics of Electron Transfer in Molecular Loop Structures: Spontaneous Breaking of Chiral Symmetry under Strong Decoherence}
\author{Nguyen Thanh Phuc}
\author{Akihito Ishizaki}
\affiliation{Department of Theoretical and Computational Molecular Science, Institute for Molecular Science, Okazaki 444-8585, Japan}
\affiliation{Department of Structural Molecular Science, The Graduate University for Advanced Studies, Okazaki 444-8585, Japan}

\date{\today}

%%%%%%%%%%%%%%%%%%%%%%%%%%
\begin{abstract}
Manipulation of quantum systems is the basis for many promising quantum technologies. However, how quantum mechanical principles can be used to manipulate the dynamics of quantum dissipative systems remains unanswered because of strong decoherence effects arising from interaction with the surrounding environment. In this work, we demonstrate that electron transfer dynamics in molecular loop structures can be manipulated with the use of Floquet engineering by applying a laser field. Despite strong dephasing, the system's dynamics spontaneously breaks the chiral symmetry of the loop in a controllable fashion, followed by the generation of a robust steady-state electronic current without an external voltage. A novel exponential scaling law that relates the magnitude of the current to the system-environment coupling strength is revealed numerically. The breaking of chiral symmetry and the consequent controllable unidirectional flow of electrons could be employed to construct functional molecular electronic circuits.
\end{abstract}

\maketitle

%%%%%%%%%%%%%%%
\section{Introduction}
\label{sec: Introduction}
%\textit{Introduction.}
Quantum manipulation of different degrees of freedom in various types of physical systems has recently attracted growing attention as an indispensable ingredient at the heart of the quantum revolution. 
Examples include the manipulation of numerous sorts of qubit platforms such as photons, trapped ions, and superconducting qubits for quantum computation and quantum communication~\cite{Nielsen-book, Ladd10, Albash18, Ekert14, Bernstein17}, the utilization of highly controllable and precisely measurable systems of ultracold atoms and molecules for quantum simulation~\cite{Bloch12, Bohn17, Moses16}, and the exploitation of unique features of systems of nitrogen-vacancy defect centers in diamond as well as of exotic quantum states such as the spin-squeezing state and entangled states for quantum sensing and quantum metrology purposes~\cite{Giovannetti11, Degen17, Pezze18}. 
As a tool of quantum manipulation, Floquet engineering has been employed to investigate various aspects of nonequilibrium dynamics in well-isolated quantum systems such as ultracold atoms~\cite{Eckardt17, Goldman14, Goldman15, Bukov15, Holthaus16, Moessner17}. In Floquet engineering, a system parameter is temporally modulated in a periodic manner.
Floquet engineering has been employed to control the superfluid-Mott-insulator phase transition in an atomic cloud of $^{87}$Rb~\cite{Zenesini09}, to simulate frustrated classical magnetism in triangular optical lattices~\cite{Struck11}, to generate artificial magnetic fields for charge-neutral particles~\cite{Aidelsburger11, Struck12, Aidelsburger13, Miyake13}, and to realize topologically nontrivial band structures~\cite{Jotzu14}. However, because quantum mechanical effects are often vulnerable to decoherence originating from interaction with the numerous dynamic degrees of freedom in the surrounding environment, it has been believed that nearly perfect isolation of the system from the environment, e.g., ultracold atomic gases, is a prerequisite for quantum manipulations such as Floquet engineering~\cite{Eckardt17}. 
Unlike well-isolated quantum systems, molecules in condensed phases are often embedded in a high density of environmental particles, leading to moderate-to-strong system-environment coupling. Consequently, the question of how quantum mechanical principles can be harnessed to manipulate the dynamics of condensed-phase molecular systems is nontrivial. Nevertheless, the present authors have recently demonstrated that Floquet engineering can significantly accelerate electronic excitation transfer between two molecules~\cite{Phuc18}. The key is that time periodic modulation of the Franck-Condon transition energy of photoactive molecules leads to minimization of the decoherence effect in a similar manner to the decoherence-free subspace~\cite{Lidar1998}. A similar mechanism can be applied to other important dynamical processes including charge and spin transfers (e.g., the spin-singlet fission) in molecular systems. In this work, we show that Floquet engineering can be employed to manipulate not only the amplitude but also the phase associated with inter-site coupling, namely the Peierls phase, which is irrelevant in the case of two-site systems~\cite{Phuc18} but can dramatically influence electron transfer (ET) in molecular networks.

Investigations of quantum dynamics in various types of network structures~\cite{Walschaers16} are significant in understanding fundamental processes such as energy and charge transfers in chemical and biological systems~\cite{Scholes:2017}, the physics of quantum-walk-related phenomena~\cite{Kempe03}, and practical applications using artificial materials such as photovoltaics~\cite{Bredas17, Green17} and photonic circuits~\cite{Crespi13, Metcalf13}. However, in the regime of moderate-to-strong system-environment coupling, quantum coherence, i.e., relative phases between electronic states at different sites in the network, can be rapidly destroyed. 
It is also noteworthy that despite quantum transport properties of various types of driven dissipative systems~\cite{Grifoni98, Gammaitoni98, Kohler05} including two-level systems~\cite{Grifoni95}, periodic tight-binding systems~\cite{Hartmann97}, and double-well potentials~\cite{Dittrich93} having been theoretically studied, up to date no attention has been paid to the control of quantum transport in periodically driven network structures via the manipulation of Peierls phase of inter-site couplings, especially in strong dissipative systems because of the effect of decoherence. 
Therefore, we need a new mechanism to protect effects of quantum manipulation on driven quantum transport in network structures under such strong decoherence, which is shown to be possible by exploiting the topological loop structure in molecular systems. 
In particular, we show that ET dynamics spontaneously break the chiral symmetry of the loop in a controllable fashion and in turn generate a robust steady-state electronic current without an external voltage that remains non-vanishing even when the system-environment interaction is stronger than the characteristic energy scale of the system. By numerically investigating the dependence of the magnitude of the current on the system-environment coupling strength, we derive a new exponential scaling law relating these two quantities. 

%%%%%%%%%%%%
%\section{Theory}
\section{Electron transfer in molecular networks}
\label{sec: Electron transfer in molecular networks}
\begin{figure}[htbp]
  \centering
  \includegraphics[keepaspectratio,width=1.8in]{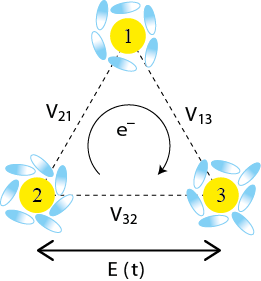}
  \caption{%\textbf{EET in molecular loop structures.} 
  Schematic illustration of ET in a molecular loop structure consisting of three sites (yellow). Each site is coupled to an independent environment illustrated by an ensemble of ellipses (blue). Electron transfer are allowed between all pairs of neighboring sites. The ET between sites $m$ and $n$ is characterized by the inter-site coupling, $V_{mn}$ ($m,n=1,2,3$). The electronic energy experiences a site-dependent time-periodic modulation caused by the oscillating electric field $\mathbf{E}(t)$ of an applied laser pulse that is aligned along the direction connecting the second and the third molecules.}
  \label{fig: system}
\end{figure}

%\textit{ET in molecular networks.}
To demonstrate the manipulation of ET dynamics in molecular networks using Floquet engineering, we consider ET in a triangular loop made of three sites as illustrated in Fig.~\ref{fig: system}. Examples of molecular loop structures include ring-shaped cyclic compounds~\cite{Ernzerhof06, Rai11, Rai12, Craven17}, donor-acceptor triads~\cite{Herranz00, Larsen18}, and the spatial arrangement of bacteriochlorophyll $a$ molecules in the light-harvesting 2 (LH2) complex of purple bacteria~\cite{McDermott95}. 
The molecular loop structure possesses a chiral symmetry, i.e., there is equality between transports in the clockwise and anti-clockwise directions.

The Hamiltonian to describe ET dynamics in condensed phases can be expressed as $\hat{H}=\hat{H}^\mathrm{el}+\hat{H}^\mathrm{env}+\hat{H}^\mathrm{int}$ with the Hamiltonian of the ET given by~\cite{Renger-Marcus:2003}
\begin{align}
    \hat{H}^\mathrm{el}=\sum_{m=1}^3 E_m|m\rangle\langle m|+\sum_{m,n=1}^3 \hbar V_{mn}|n\rangle\langle m|.
    \label{eq: original Hamiltonian}
\end{align}
In the above, 
$\lvert m \rangle$ denotes the state where only the $m$th molecule is reduced and negatively charged while the other molecules are in the  electronically ground neutral states, i.e. an excess electron is located on the $m$th site.
The energy of state $\lvert m \rangle$ is given by $E_m$ when the energy of the ``vacuum state'' where all the molecules are in the 
electronically ground neutral states is set to be zero.
The real number $\hbar V_{mn}$ denotes the inter-site coupling to drive ET reaction between the $m$th and $n$th molecules.

The Hamiltonian of the environment associated with the $m$th site is given by $\hat{H}_m^\mathrm{env}=\sum_\xi\hbar\omega_{m,\xi} \hat{b}_{m,\xi}^\dagger\hat{b}_{m,\xi}$, where $\hat{b}_{m,\xi}$ denotes the annihilation operator of the $\xi$th mode of the environment with frequency $\omega_{m,\xi}$. The electronic energy at the $m$th site experiences fluctuations caused by the environmental dynamics as expressed by the interaction Hamiltonian 
$\hat{H}^\mathrm{int}=|m\rangle\langle m|\sum_\xi g_{m,\xi} (\hat{b}_{m,\xi}^\dagger+\hat{b}_{m,\xi} )$, 
where $g_{m,\xi}$ denotes the coupling strength of the $\xi$th mode. This form of the interaction Hamiltonian induces environmental reorganization. When an excess electron is localized on a molecule, the environment associated with that molecule would make a change in its configuration to a new equilibrium position that is shifted from its equilibrium position when the molecule is in the electronically ground neutral state. The environmental reorganization and its timescale are characterized by the relaxation function $\Psi_m(t)=(2/\pi)\int_0^\infty \text{d}\omega J_m(\omega)\cos(\omega t)/\omega$, 
where $J_m(\omega)$ stands for the spectral density 
$J_m(\omega)=\pi\sum_\xi g_{m,\xi}^2\delta(\omega-\omega_{m,\xi})$
of the environment.
Generally, the relaxation function may have a complex form involving various components. For the sake of simplicity, however, we model the function using an exponential form, $\Psi_m(t)=2\lambda_m \exp(-t/\tau_m)$, where $\lambda_m$ is the environmental reorganization energy, which is usually employed to characterize the system-environment coupling strength, and $\tau_m$ is the characteristic timescale of the environmental relaxation or reorganization process \cite{Ishizaki10}.

In the following, we consider the case of independent environments associated with three sites \cite{Kato:2018jw}. In general, there can be some correlation between fluctuations caused by environments associated with different sites in a molecular network as the change of configuration of the environment around one molecule as ET occurs can affect the configurations of environments of other molecules \cite{Weiss:4th,Fujihashi:2016iga}. 
As shown in Appendix~\ref{sec: ET dynamics in the presence of environments with correlated/anticorrelated fluctuations}, however, the main results of this paper remain unchanged, at least qualitatively, for both cases of correlated and anticorrelated fluctuations.
For the sake of simplicity, we set $\tau_m = \tau$, $\lambda_m = \lambda$, $\Omega_m^0=\Omega^0$, and $V_{mn}=V$ for $m,n=1,2,3$.

%%%%%%%%%%%%%%%%%%
\section{Floquet engineering}
\label{sec: Floquet engineering}
%\textit{Floquet engineering.}
In Floquet engineering, the Hamiltonian of the system is temporally modulated in a periodic manner by, for example, applying a laser pulse through the Stark effect $E_\mathrm{S}(t)=-\mathbf{E}(t)\cdot\boldsymbol{\mu}$, where $\mathbf{E}(t)$ and $\boldsymbol{\mu}$ are the oscillating electric field and the electric dipole moment, respectively. 
The time-dependent electronic energy at the $m$-th site is expressed as $E_m(t)=E_m^0+\hbar u_m(t)$, where the modulation $u_m(t)$ needs to satisfy $u_m(t)=u_m(t+T)$ and $\int_0^T d t\,u_m(t)=0$ with  $T$ being the modulation period. In the limit of high driving frequency, the system's dynamics are characterized by an effective time-independent Hamiltonian $\hat{H}_\mathrm{eff}$ obtained by taking the lowest-order terms in the high-frequency expansion \cite{Eckardt17}. This has the same form as the original Hamiltonian in Eq.~\eqref{eq: original Hamiltonian}; however, the inter-site coupling $V_{mn}$ is replaced by 
    $V_{mn}^\mathrm{eff} 
    =
    (V_{mn}/T)
    \int_0^T d t\, 
    \exp\left\{ i[\chi_n(t)-\chi_m(t)] \right\}$ 
with $\chi_m(t)=\int_0^t d t'\, u_m(t')$~\cite{Phuc18}. 

In general, the effective inter-site coupling, $V_{mn}^\mathrm{eff}$, is a complex number expressed as 
    $V_{mn}^\mathrm{eff}
    =
    \lvert V_{mn}^\mathrm{eff} \rvert 
    \exp({i\,\theta_{mn}^\mathrm{eff}})$,
where $\theta_{mn}^\mathrm{eff}$ is termed the effective Peierls phase.
For a sinusoidal modulation, $V_{mn}^\mathrm{eff}$ takes real values~\cite{Phuc18} and thus $\theta_{mn}^\mathrm{eff}$ takes only the value of 0 or $\pi$. 
However, it was shown that the effective inter-site coupling can take a complex value, provided that the time-periodic modulation breaks two special symmetries: reflection symmetry for a suitable time $\tau$, namely $u_m(t-\tau)=u_m(-t-\tau)$, and shift antisymmetry $u_m(t-T/2)=-u_m(t)$~\cite{Struck12}. 
For example, we can take a time-periodic modulation composed of two sinusoids, $u_m(t)=A_m\sin(\omega t)+B_m\sin(2\omega t)$, by applying a shaped laser pulse~\cite{Weiner00}. Since the wavelength of the laser is typically much larger than the size of the molecular system, the electric field of the laser can be considered as homogeneous over the whole system. 
If the electric field of the linearly polarized laser is aligned along the direction connecting the second and the third sites, the differences in energy modulations between the three sites are
 $A_1-A_2=A_3-A_1=(A_3-A_2)/2\equiv A/2$ and $B_1-B_2=B_3-B_1=(B_3-B_2)/2\equiv B/2$, where the driving amplitudes $A$ and $B$ are approximately proportional to the magnitude of the electric field, the charge of electron, and the distance between neighboring sites. 
The ET dynamics in the closed loop depends on the effective total Peierls phase $\theta_\mathrm{tot}^\mathrm{eff}=\theta_{13}^\mathrm{eff}+\theta_{32}^\mathrm{eff}+\theta_{21}^\mathrm{eff}$ accumulated by the electron as it travels in the clockwise direction. The dependence of $\theta_\mathrm{tot}^\mathrm{eff}$ on the driving amplitudes $A$ and $B$ is given in details in Appendix~\ref{sec: Effective electron transfer matrix elements for a driving composed of two sinusoids}.

In the following numerical demonstrations, since the energy fluctuation is proportional to the temperature, we consider $\hbar V\lesssim k_\mathrm{B}T$ for the strong decoherence regime. Here, we set $T=300\,{\rm K}$ and $V=50$ cm$^{-1}$, which is of the same order of magnitude as the ET in photosynthetic reaction center~\cite{Novoderezhkin11}.
The environmental relaxation time $\tau$ is also taken to be a typical value, $\tau=100\,{\rm fs}$. 
The driving frequency $\omega$ is taken to be $\omega=200\,{\rm THz}$. 
For describing ET dynamics in the molecular network, an adequate description is provided with the reduced density operator $\hat\rho(t)$, i.e., the partial trace of the density operator of the total system over the environmental degrees of freedom. 
The time evolution of the reduced density operator with the initial condition of $\hat{\rho}(t=0)=\lvert 1 \rangle\langle 1 \rvert$ can be solved in a numerically accurate fashion through the use of, for example, the hierarchical equations of motion approach~\cite{Tanimura06}. Technical details are given in Ref.~\cite{Ishizaki10}.

%%%%%%%%%%%%%%%%
%\section{Results and Discussion}
%\label{sec: Results and Discussion}

%%%%%%%%%%%%%%%%%%
\section{Spontaneous breaking of chiral symmetry}
\label{sec: Spontaneous breaking of chiral symmetry}
\begin{figure}[tbp]
  \centering
  \includegraphics[keepaspectratio]{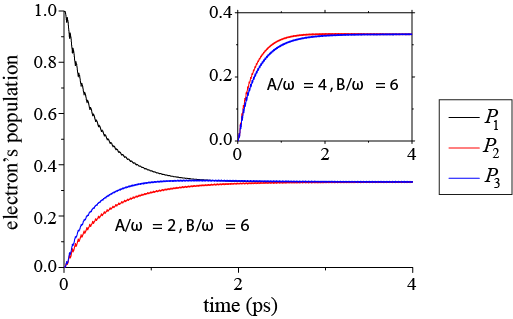}
  \caption{Time evolutions of the electronic populations $P_m$ ($m=1,2,3$) at three sites in the molecular loop structure under Floquet engineering. The system is driven by a laser pulse composed of two sinusoids with frequencies $\omega$ and $2\omega$. The two corresponding driving amplitudes are $A/\omega=2$ and $B/\omega=6$. Inset: Time evolutions of $P_2$ (red) and $P_3$ (blue) for $A/\omega=4$. $B$ is kept constant.}
  \label{fig: time evolution of site population with a periodic driving}
\end{figure}

%\textit{Spontaneous breaking of chiral symmetry.}
We consider the case where the reorganization energy is comparable to the inter-site coupling, $\lambda=V$, and the system is driven with the driving amplitudes $A/\omega=2$ and $B/\omega=6$.
It is noteworthy that the dynamics of the system under consideration is in a non-Markovian regime as both the system-environment coupling strength and the environment's relaxation time scale are comparable in magnitude with the characteristic energy scale of the system. As a result, the memory effect of the environment must be taken into account as opposed to the Floquet-Markov master equation that has been widely used in studying driven quantum transport~\cite{Kohler05}.
Time evolutions of the electronic populations at three sites are shown in Fig.~\ref{fig: time evolution of site population with a periodic driving}. The populations rapidly equilibrates to the steady-state values $P_1(t\to\infty) = P_2(t\to\infty) = P_3(t\to\infty) = 1/3$, showing no noticeable signature of quantum interference due to the strong decoherence. 
Nevertheless, there is clearly a population imbalance between the second and the third sites, $P_2(t)<P_3(t)$, over a relatively long time period of approximately 2 ps. It should be noted that the three sites are identical, leading to a chiral symmetry between two possible ET pathways along the loop in the clockwise and anti-clockwise directions. Meanwhile, the AC driving also preserves this symmetry on time average, which is relevant to the high-driving-frequency limit under consideration. It should also be noted that the applied laser field under consideration is linearly polarized, in contrast to the case of circularly polarized field used to generate ring currents~\cite{Barth06, Barth06b, Nobusada07} which explicitly breaks the chiral symmetry.
Therefore, the emergent imbalance in the population distribution of electron between the two sites implies that ET dynamics under Floquet engineering spontaneously breaks the chiral symmetry of the molecular loop structure. 

Moreover, the direction of the emergent chirality can be manipulated by varying the driving amplitudes. It is evident from the inset of Fig.~\ref{fig: time evolution of site population with a periodic driving} that the relative electronic distributions at the two sites are reversed when the driving amplitude varies from $A/\omega=2$ to $A/\omega=4$ with keeping $B$ constant, indicating that the emergent chirality of the system can be fully controlled by Floquet engineering. It should be noted that the direction of the emergent chirality changes by only varying the driving amplitude without any change of phase or direction or other parameters of the modulation, reflecting the fact that the chiral symmetry is spontaneously broken by the ET dynamics under Floquet engineering.
The controlling of the emergent chirality can be understood through the sign of the effective total Peierls phase. Indeed, $\theta_\mathrm{tot}^\mathrm{eff}$ changes its sign from negative to positive when the driving amplitude changes from $A/\omega=2$ to $A/\omega=4$ (see Appendix~\ref{sec: Role of the effective total Peierls phase in the EET dynamics in molecular loop structures} for details).

 The small zigzag pattern on top of the time evolution in Fig~\ref{fig: time evolution of site population with a periodic driving}, usually called micromotion, is a direct consequence of a time-periodic driving with finite frequency. It corresponds to higher-order terms in the high-frequency expansion; therefore, the higher the driving frequency, the smoother the time evolution. In the following sections, to remove the micromotion and in turn obtain smooth time evolution, we consider the high-driving-frequency limit in which ET dynamics are characterized by the effective inter-site coupling $V_{mn}^\mathrm{eff}$. Experimentally the micromotion can be removed by filtering out high-frequency components from the measured signal.

%%%%%%%%%%%%%%%%%%
\section{Robust electronic current without an external voltage}
\label{sec: Electronic current without an external voltage}
%\textit{Electronic current.}
It is not only the electronic population but also the electronic current flowing in the molecular loop that are substantially affected by Floquet engineering. The current is defined as $I_{mn}(t)=-e\mathrm{Tr}[ \hat{I}_{mn}\hat{\rho}(t) ]$ ($m,n=1,2,3$), where $-e$ is the unit charge of an electron and $\hat{I}_{mn}=-i\left(V_{mn} \lvert n \rangle\langle m \rvert -V_{mn}^\ast \lvert m\rangle\langle n \rvert \right)$ is the current operator describing the flow of electron from the $m$th to the $n$th sites. 
The time evolutions of the current $I_{32}(t)$ for the two different driving amplitudes $A/\omega=2$ and $A/\omega=4$ considered above are shown in Fig.~\ref{fig: current versus Peirels phase}, where $B$ is kept constant. It is evident that $I_{32}(t)$ makes a sharp rise to a positive (negative) value at early time, then undergoes a short damped oscillation before approaching the steady-state value which is positive (negative) for $A/\omega=2$ ($A/\omega=4$). The non-zero electronic current between the two molecules reflects the breaking of chiral symmetry while the dependence of the direction of current $I_{32}$ both at early time and in the steady state on the driving amplitude indicates again that the emergent chirality can be fully controlled under Floquet engineering. Since the electron's populations $P_m(t)$ ($m=1,2,3$) are time-independent in the steady state, it must be that $I_{13}(t\to\infty)=I_{32}(t\to\infty)=I_{21}(t\to\infty)\equiv I_\mathrm{ss}$.   

\begin{figure}[tbp]
  \centering
  \includegraphics[keepaspectratio]{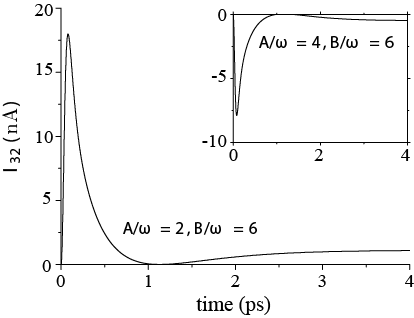}
  \caption{Time evolution of the electronic current $I_{32}$ flowing from the third to the second site in the molecular loop structure under Floquet engineering. The driving amplitudes are $A/\omega=2$ and $B/\omega=6$. Inset: Time evolution of $I_{23}$ for $A/\omega=4$, where $B$ is kept constant.}
  \label{fig: current versus Peirels phase}
\end{figure}

\begin{figure}[tbp]
  \centering
  \includegraphics[keepaspectratio]{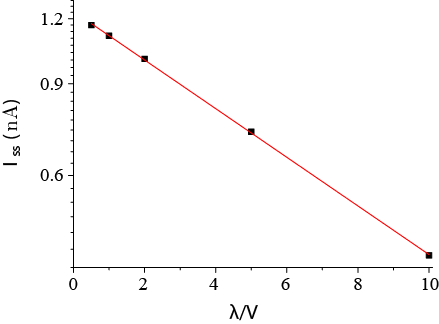}
  \caption{Dependence of the magnitude of the steady-state electronic current $I_\mathrm{ss}$ flowing in the molecular loop structure on the reorganization energy $\lambda$ characterizing the system-environment coupling strength and normalized by the inter-site coupling $V$. The vertical axis is displayed using a logarithmic scale. The red line demonstrates the linear fitting.}
  \label{fig: current vs reorganization energy}
\end{figure}

%%%%%%%%%%%%%%%%%%%%%
%\textit{Scaling law of the steady-state current.}
Despite the strong effect of decoherence, the steady-state electronic current is non-vanishing.
To investigate the effect of decoherence on the steady-state current, we repeated the calculation of $I_\mathrm{ss}$ for different values of the environmental reorganization energy $\lambda$ with the driving amplitudes fixed to be $A/\omega=2$ and $B/\omega=6$. The dependence of $I_\mathrm{ss}$ on $\lambda$ is plotted in Fig.~\ref{fig: current vs reorganization energy} with a logarithmic scale on the vertical axis. The clearly observed linear correlation in the figure indicates an exponential dependence $I_\mathrm{ss}=I_0 e^{-\kappa (\lambda/V)}$ between the steady-state current and the system-environment coupling strength with the constants $I_0\simeq 1.24\, {\rm nA}$ and $\kappa\simeq 0.11$ obtained numerically by using a standard procedure of linear fitting. It should be noted that, while an exponential time decay of quantum coherence has often been derived in typical models of dephasing, here we found a different exponential scaling law for the effect of decoherence in terms of a steady-state current that persists as long as the modulation is applied. 

It is understood from the small damping factor $\kappa\ll1$ of the steady-state current that it remains non-vanishing even for a very strong system-environment coupling where the environmental reorganization energy is much larger than the characteristic energy scale of the system's dynamics. This is unlike the normal situation in open quantum systems, where typical physical quantities often decay very quickly owing to the strong effect of decoherence.  The robustness of the steady-state current even in the presence of very strong system-environment coupling can be attributed to a classical topological feature of the effect that drives the current. 
Indeed, the effect of a nonzero effective total Peierls phase in a molecular loop structure is equivalent to the effect of a magnetic field on a charged particle moving in closed loop as shown by the Aharonov-Bohm effect~\cite{Aharonov59}. Since the effect of a magnetic field on a charged particle  persists in the classical regime, it is reasonable to expect that the effect is robust against decoherence. 

%%%%%%%%%%%%%%%%%%%%
\section{Conclusion}
\label{sec: Conclusion}
%\textit{Conclusion.}
We have investigated how the quantum dynamics of ET in molecular loop structures can be manipulated using Floquet engineering. 
We found that, despite strong dephasing, ET dynamics can spontaneously break the chiral symmetry of the loop in a controllable fashion and in turn generate a robust steady-state electronic current without an external voltage that remains non-vanishing even in the strong system-environment-coupling limit. Employing this kind of topological loop structure in quantum manipulation for protection from environmental effects can serve as a useful guide for controlling dynamical processes in molecular systems.
 A new exponential scaling law that relates the magnitude of the steady-state current to the system-environment coupling strength was also revealed numerically. 
 
 The controllable unidirectional flow of electrons following the breaking of chiral symmetry in molecular loop structures can find numerous applications in, for example, constructing functional molecular electronic circuits. 
 The steady-state electronic current generated in molecular loop structures without an external voltage is closely related to the persistent current in mesoscopic normal metal rings observed at very low temperature ($\lesssim 1$ K) in the presence of an external magnetic field~\cite{Buttiker83, Bluhm09, Jayich09}, although in which case the effect of decoherence is dominated by that of static disorders. However, compared with mesoscopic normal metal rings, an electronic current of the same order of magnitude or even higher can be generated in nanoscale molecular loop structures at room temperature. 

It is also noteworthy that controlling electron transfer in molecular loop structures by a static magnetic field through the Aharonov-Bohm effect (ABE) is experimentally challenging because the required field strength is unrealistically high, of the order of $10^4$ T~\cite{Walczak04, Maiti07} or various kinds of conditions~\cite{Rai11, Rai12} including small dephasing are needed. Therefore, the control of electron dynamics in molecular systems with the use of Floquet engineering via the Peirels phase, which works even in the presence of strong decoherence, can open a new route towards constructing ABE-based molecular circuits.
Although this paper focuses on the study of ET dynamics, similar results should be expected for other important dynamical processes such as electronic excitation transfer in condensed-phase molecular systems if we can make a site-dependent time-periodic modulation of the Franck-Condon transition energy in the molecular network.  

%%%%%%%%%%%%%%%%%%%%%%%%%%
\begin{acknowledgements}
N.T.P. thanks Yuta~Fujihashi for his assistance in preparing the figures.
This work was supported by JSPS KAKENHI Grant Numbers 17H02946 and 18H01937 as well as JSPS KAKENHI Grant Number 17H06437 in Innovative Areas "Innovations for Light-Energy Conversion (I$^4$LEC)."
\end{acknowledgements}

%%%%%%%%%%%%%%%%%%
\begin{appendix}
%%%%%%%%%%%%%%%%%%%%%%%%%%
\section{ET dynamics in the presence of environments with correlated/anticorrelated fluctuations}
\label{sec: ET dynamics in the presence of environments with correlated/anticorrelated fluctuations}
In the main text, we considered the case of independent environments associated with three sites. In general, there can be some correlation between fluctuations caused by environments associated with different sites in a molecular network as the change of configuration of the environment around one molecule as ET occurs can affect the configurations of environments of other molecules. In this section, we consider two cases in which the energy fluctuations caused by environments around neighboring molecules have a perfect correlation/anticorrelation~\cite{Ishizaki10}. 

The correlation/anticorrelation between energy fluctuations caused by environments associated with neighboring sites is taken into account by changing the operators of the system that couple to the environments from $\hat{V}_{m}=|m\rangle\langle m|$ ($m=1,2,3$) to 
\begin{align}
\hat{C}_{m}=\frac{1}{\sqrt{2}}\left(|m\rangle\langle m|\pm |m+1\rangle\langle m+1|\right),
\end{align}
where $m=1,2,3$ ($m+1$ would be 1 if $m=3$) and $\pm$ corresponds to correlation/anticorrelation. Here the normalization factor $1/\sqrt{2}$ is introduced so that the total reorganization energy of the environment at each site is the same as in the case of independent environments. We perform numerically accurate quantum dynamics calculations of the ET dynamics in a way similar to the case of independent environments. 

\begin{figure}[tbp] % float placement: (h)ere, page (t)op, page (b)ottom, other (p)age
  \centering
  % file name: G:/Draft-Control of quantum dynamics in condensed-phase molecular loop (July 2018)/Figures/correlated.eps
  \includegraphics[keepaspectratio]{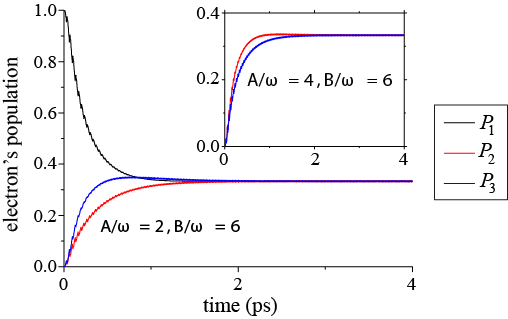}
  \caption{Time evolutions of the electron's populations $P_m$ ($m=1,2,3$) at three sites in the molecular loop structure under Floquet engineering with correlated energy fluctuations caused by the environments. The system is driven by a laser pulse composed of two sinusoids with frequencies $\omega$ and $2\omega$. The two corresponding driving amplitudes are $A/\omega=2$ and $B/\omega=6$. Inset: Time evolutions of $P_2$ (red) and $P_3$ (blue) for $A/\omega=4$ ($B$ is kept constant).}
  \label{fig: correlated}
\end{figure}

\begin{figure}[tbp] % float placement: (h)ere, page (t)op, page (b)ottom, other (p)age
  \centering
  % file name: G:/Draft-Control of quantum dynamics in condensed-phase molecular loop (July 2018)/Figures/anticorrelated.eps
  \includegraphics[keepaspectratio]{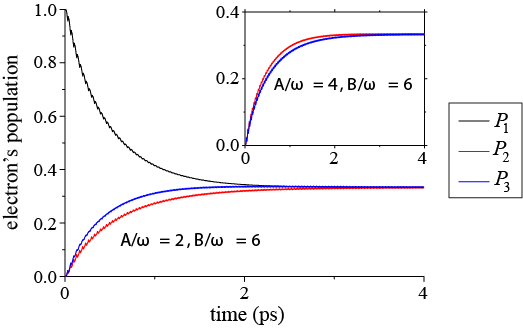}
  \caption{Time evolutions of the electron's populations $P_m$ ($m=1,2,3$) at three sites in the molecular loop structure under Floquet engineering with anticorrelated energy fluctuations caused by the environments. The parameters are the same as those in Fig.~\ref{fig: correlated}.}
  \label{fig: anticorrelated}
\end{figure}

The time evolutions of the electron's populations at three sites are shown in Fig.~\ref{fig: correlated} (Fig.~\ref{fig: anticorrelated}) for the case of correlated (anticorrelated) fluctuations. It is clear that there is a population imbalance of electron between the second and the third sites, reflecting the spontaneous breaking of the chiral symmetry of the molecular loop structure, as in the case of independent environments. Moreover, the direction of the emergent chirality can also be fully controlled by varying the driving amplitudes through the change of the effective total Peirels phase. Therefore, it should be expected that the main results in the main text remain unchanged even if correlation in fluctuations caused by different environments exists.

%%%%%%%%%%%%%%%%%%%%%%%%%%
\section{Effective inter-site coupling for a driving composed of two sinusoids}
\label{sec: Effective electron transfer matrix elements for a driving composed of two sinusoids}
In Floquet engineering, the Hamiltonian of the system is temporally modulated in a periodic manner by, for example, applying a laser pulse. The time-dependent electronic energy at the $m$-th site is expressed as $E_m(t)=E_m^0+u_m(t)$, where the modulation $u_m(t)$ needs to satisfy $u_m(t)=u_m(t+T)$ and $\int_0^T d t\,u_m(t)=0$ with  $T$ being the modulation period. In the limit of high driving frequency, the system's dynamics are characterized by an effective time-independent Hamiltonian $\hat{H}_\mathrm{eff}$ obtained by taking the lowest-order terms in the high-frequency expansion \cite{Eckardt17}. This has the same form as the original Hamiltonian; however, the inter-site coupling $V_{mn}$ is replaced by 
    $V_{mn}^\mathrm{eff} 
    =
    (V_{mn}/T)
    \int_0^T d t\, 
    \exp\left\{ i[\chi_n(t)-\chi_m(t)] \right\}$ 
with $\chi_m(t)=\int_0^t d t'\, u_m(t')$~\cite{Phuc18}. 

If the time-periodic modulation is composed of two sinusoids, $u_m(t)=A_m\sin(\omega t)+B_m\sin(2\omega t)$, we have
\begin{align}
\chi_m(t)-\chi_n(t)=&\left(\frac{A_m-A_n}{\omega}\right)(1-\cos\omega t)\nonumber\\
&+\left(\frac{B_m-B_n}{2\omega}\right)(1-\cos2\omega t).
\end{align}
The effective inter-site coupling is then given by
\begin{align}
\frac{V_{mn}^\mathrm{eff}}{V_{mn}}=&\frac{1}{T} \int_0^T dt\,
\exp \left\{i\left[\left(\frac{A_m-A_n}{\omega}\right)\cos\omega t \right.\right. \nonumber\\
&\left.\left.+\left(\frac{B_m-B_n}{2\omega}\right)\cos2\omega t
\right]\right\}.
\end{align}
Here, the time-independent factors of $\exp\left(\frac{A_n-A_m}{\omega}\right)$ and $\exp\left(\frac{B_n-B_m}{2\omega}\right)$ will be canceled out when the electron travels over one loop.

By setting $x\equiv \omega t$, we need to evaluate the integral
\begin{align}
\frac{V_{mn}^\mathrm{eff}}{V_{mn}}=&\frac{1}{2\pi} \int_0^{2\pi} dx\,
\exp\left\{i\left[\left(\frac{A_m-A_n}{\omega}\right) \cos x\right.\right.\nonumber\\
&\left.\left.+\left(\frac{B_m-B_n}{2\omega}\right)\cos 2x\right]\right\} \nonumber\\
\equiv& f\left(\frac{A_m-A_n}{\omega},\frac{B_m-B_n}{2\omega}\right).
\end{align}
If the laser's electric field is aligned along the direction connecting the second and the third sites, the differences between electronic energies at three sites are
\begin{align}
A_1-A_2=A_3-A_1=(A_3-A_2)/2\equiv A/2,\\
B_1-B_2=B_3-B_1=(B_3-B_2)/2\equiv B/2.
\end{align}
The effective total Peierls phase $\theta_\mathrm{tot}^\mathrm{eff}=\theta_{13}^\mathrm{eff}+\theta_{32}^\mathrm{eff}+\theta_{21}^\mathrm{eff}$ is then given by 
\begin{align}
\theta_\mathrm{tot}^\mathrm{eff}=\text{Arg}\left[f\left(\frac{A}{\omega},\frac{B}{2\omega}\right)\right]
-2\text{Arg}\left[f\left(\frac{A}{2\omega},\frac{B}{4\omega}\right)\right],
\end{align}
where $\text{Arg}[x]$ is the argument of a complex number $x$.

\begin{figure}[tbp] % float placement: (h)ere, page (t)op, page (b)ottom, other (p)age
  \centering
  % file name: G:/Draft-Control of quantum dynamics in condensed-phase molecular loop (July 2018)/Figures/dbl_freq_Peirels.eps
  \includegraphics[keepaspectratio]{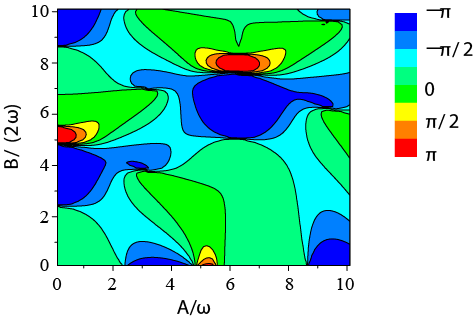}
  \caption{Effective total Peierls phase $\theta_\mathrm{tot}^\mathrm{eff}$ as a function of the driving amplitudes $A/\omega$ and $B/(2\omega)$.}
  \label{fig: double frequency Peirels phase}
\end{figure}

\begin{figure}[tbp] % float placement: (h)ere, page (t)op, page (b)ottom, other (p)age
  \centering
  % file name: G:/Draft-Control of quantum dynamics in condensed-phase molecular loop (July 2018)/Figures/dbl_freq_amp.eps
  \includegraphics[keepaspectratio]{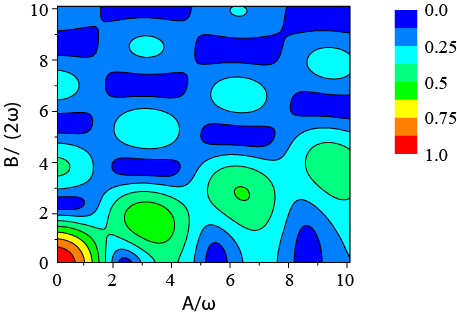}
  \caption{Magnitude of the effective inter-site coupling $\left|V_{32}^\mathrm{eff}/V_{32}\right|$ as a function of the driving amplitudes $A/\omega$ and $B/(2\omega)$.}
  \label{fig: double frequency amplitude}
\end{figure}

The effective total Peierls phase $\theta_\mathrm{tot}^\mathrm{eff}$ and the magnitude of the effective inter-site coupling $\left|V_{32}^\mathrm{eff}/V_{32}\right|$ as functions of the driving amplitudes $A/\omega$ and $B/(2\omega)$ are shown in Figs.~\ref{fig: double frequency Peirels phase} and \ref{fig: double frequency amplitude}, respectively.

%%%%%%%%%%%%%%%%%%%%%%%%
\section{Role of the effective total Peierls phase in the ET dynamics in molecular loop structures}
\label{sec: Role of the effective total Peierls phase in the EET dynamics in molecular loop structures}
In order to get insights into the role of the effective total Peierls phase in the ET dynamics in molecular loop structures, in this section we will temporarily neglect the dynamical coupling of the system to the environments and consider only a canonical distribution of energy eigenstates of the system at temperature $T$ under a nonzero effective total Peierls phase. For a single electron, there are a total of three energy eigenstates $|\psi_j\rangle$ ($j=1,2,3$) with the corresponding energy eigenvalues $E_j$. These eigenstates and eigenvalues can be obtained by diagonalizing the effective Hamiltonian of the system
\begin{align}
H_\mathrm{s}^\mathrm{eff}=\sum_{m,n=1}^3\hbar V_{mn}^\mathrm{eff}|n\rangle\langle m|,
\label{eq: Hamiltonian}
\end{align}
where $V_{mn}^\mathrm{eff}=(V_{nm}^\mathrm{eff})^*$ and $|m\rangle$ represents the state where the electron is located at the $m$-th site. The electronic current flowing from the $m$th to the $n$th site is given by 
\begin{align}
I_{mn}(t)=\text{Tr}[\hat{I}_{mn}\hat{\rho}(t)],
\label{eq: current formula}
\end{align}
where $\hat{I}_{mn}=-i(V_{mn}|n\rangle\langle m|-V_{mn}^*|m\rangle\langle n|)$ is the current operator. This current operator satisfies the continuity equation $\partial \hat{n}_1/\partial t=\hat{I}_{21}(t)-\hat{I}_{13}(t)$, where $\hat{n}_1$ is the number operator of electron at the first site, and similarly for $\hat{n}_2$ and $\hat{n}_3$. For the energy eigenstates, since the number of electrons at each molecule is stationary, the electronic currents flowing in the clockwise direction between different molecules must be equal, $I_{13}=I_{32}=I_{21}=I$.

First, we show that the energy eigenvalues and the magnitude of the electronic current $I$ for the corresponding energy eigenstates depend only on the effective total Peierls phase $\theta^\mathrm{eff}_\mathrm{tot}=\theta_{13}^\mathrm{eff}+\theta_{32}^\mathrm{eff}+\theta_{21}^\mathrm{eff}$ and the magnitudes of the electron's effective inter-site couplings $|V_{mn}^\mathrm{eff}|$ ($m,n=1,2,3$). 
Indeed, let us consider two sets of effective inter-site couplings $\{V_{13}, V_{32}, V_{21}\}$ and $\{V'_{13}, V'_{32}, V'_{21}\}$ with equal magnitudes, $|V_{13}|=|V_{13}'|$, $|V_{32}|=|V_{32}'|$, and $|V_{12}|=|V_{12}'|$, and equal effective total Peierls phase, $\theta_{13}+\theta_{32}+\theta_{21}=\theta_{13}'+\theta_{32}'+\theta_{21}'$. Here the superscript eff has been omitted for a simplification in notation. 
Suppose $|\psi\rangle=c_1|1\rangle+c_2|2\rangle+c_3|3\rangle$ is an energy eigenstate with eigenvalue $E$ for the set of inter-site couplings $\{V_{13}, V_{32}, V_{21}\}$. We then have 
\begin{align}
V_{21}c_2+V_{31}c_3=&Ec_1,\\
V_{12}c_1+V_{32}c_3=&Ec_2,\\
V_{13}c_1+V_{23}c_2=&Ec_3.
\end{align}
By a straightforward calculation, we can confirm that $|\psi'\rangle=c_1'|1\rangle+c_2'|2\rangle+c_3'|3\rangle$ with $c_1'=c_1$, $c_2'=c_2e^{-i\delta_{21}}$, and $c_3'=c_3e^{i\delta_{13}}$ is an energy eigenstate with the same eigenvalue $E$ for the set of inter-site couplings $\{V_{13}', V_{32}', V_{21}'\}$. Here, $\delta_{mn}=\theta_{mn}'-\theta_{mn}$ is the difference in the Peierls phase between the two sets of inter-site couplings. Moreover, for the energy eigenstate $|\psi\rangle$, the electronic current is given by $I_{mn}=2\text{Im}\{V_{mn}c_mc_n^*\}$, and similarly for $|\psi'\rangle$. Therefore, we can justify that $I_{13}=I_{13}'$, $I_{32}=I_{32}'$, and $I_{21}=I_{21}'$, i.e., the electronic current $I$ is the same for two energy eigenstates $|\psi\rangle$ and $|\psi'\rangle$. The magnitude of the electronic current for the canonical distribution of the energy eigenstates is given by 
\begin{align}
I_\mathrm{ca}=\sum_{j=1}^3 n_jI_j,
\label{eq: current for canonical distribution}
\end{align}
where $I_j$ is the current for the energy eigenstate $|\psi_j\rangle$ with eigenvalue $E_j$ and $n_j=e^{-\beta E_j}/Z$ with $\beta\equiv 1/(k_\mathrm{B}T)$ and $Z=\sum_{j=1}^3 e^{-\beta E_j}$ being the partition function. Since both $E_j$ and $I_j$ are equal for two sets of inter-site couplings $\{V_{13}, V_{32}, V_{21}\}$ and $\{V'_{13}, V'_{32}, V'_{21}\}$, the electronic current $I_\mathrm{ca}$ for the canonical distribution of energy eigenstates depends only on the effective total Peierls phase $\theta_\mathrm{tot}^\mathrm{eff}$ and the magnitudes of inter-site couplings.

Second, we show that the direction of electronic current $I_\mathrm{ca}$ for the canonical distribution of energy eigenstates depends on the sign of the effective total Peierls phase. We assume for simplicity that the three inter-site couplings have equal magnitudes, $|V_{13}|=|V_{32}|=|V_{21}|\equiv V_0$. As shown above, the current $I_\mathrm{ca}$ depends only on the effective total Peierls phase. Therefore, we can take, for example, $\theta_{13}=\theta_{21}=0$ and $\theta_{32}=\theta_\mathrm{tot}^\mathrm{eff}$. Given a value of $\theta_\mathrm{tot}^\mathrm{eff}\in (-\pi,\pi)$, the energy eigenstates and eigenvalues are obtained by numerically diagonalizing the Hamiltonian~\eqref{eq: Hamiltonian}. The electronic current for the canonical distribution of these eigenstates is then calculated by using Eqs.~\eqref{eq: current for canonical distribution} and \eqref{eq: current formula}. Here the temperature is taken to be $k_\mathrm{B}T=\hbar V_0$. The dependence of $I_\mathrm{ca}$ on $\theta_\mathrm{tot}^\mathrm{eff}$ is shown in Fig.~\ref{fig: current versus Peirels phase}. It is clearly evident that the direction of the electronic current for the canonical distribution of energy eigenstates is determined by the sign of the effective total Peierls phase. This dependence can be understood from the Aharonov-Bohm effect~\cite{Aharonov59} in which the effect of a nonzero total Peierls phase on the motion of the electron in the molecular loop structure is equivalent to that of an external magnetic field. The direction of the magnetic field is mapped to the sign of the total Peierls phase.

\begin{figure}[tbp] % float placement: (h)ere, page (t)op, page (b)ottom, other (p)age
  \centering
  % file name: E:/Draft-Control of quantum dynamics in condensed-phase molecular loop (July 2018)/Figures/current_vs_Peirelsphase_Supp_2.eps
  \includegraphics[keepaspectratio]{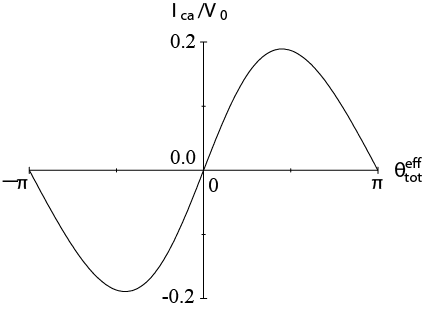}
  \caption{Electronic current $I_\mathrm{ca}$ for the canonical distribution of energy eigenstates (normalized by the inter-site coupling $V_0$) as a function of the effective total Peierls phase $\theta_\mathrm{tot}^\mathrm{eff}$. Here, $|V_{13}|=|V_{32}|=|V_{21}|\equiv V_0$ and the temperature $k_\mathrm{B}T=\hbar V_0$.}
  \label{fig: current versus Peirels phase}
\end{figure}

Following the Floquet theory~\cite{Eckardt17, Phuc18}, the effective total Peierls phase is found to be $\theta_\mathrm{tot}^\mathrm{eff}\simeq -0.41\pi$ for the driving amplitudes $A/\omega=2$ and $B/\omega=6$. Here $\omega$ is the driving frequency. The magnitudes of the effective inter-site couplings are $|V_{32}^\mathrm{eff}|\simeq 0.26V$ and $|V_{21}^\mathrm{eff}|=|V_{13}^\mathrm{eff}|\simeq 0.41V$, where $V=|V_{21}|=|V_{32}|=|V_{13}|$ is the magnitude of the bare inter-site couplings. By diagonalizing the effective Hamiltonian~\eqref{eq: Hamiltonian}, we can numerically calculate the energy eigenvalues and eigenstates in the same way as above. For $V=50$ cm$^{-1}$, the electronic current flowing in the anti-clockwise direction in the molecular loop structure for the canonical distribution of energy eigenstates at temperature $T=300$ K is found to be $I_\mathrm{ca}\simeq 1.2$ nA. As the magnitude of the steady-state current in the presence of the dynamical coupling between the system and the environments is found in the main text to be $I_\mathrm{ss}\simeq 1.12$ nA (for $\lambda=V$), it can be seen that in addition to the anti-clockwise direction of both $I_\mathrm{ca}$ and $I_\mathrm{ss}$ as expected for a negative effective total Peierls phase, the magnitudes of $I_\mathrm{ca}$ and $I_\mathrm{ss}$ are very close to each other. One reason is because the temperature under consideration $k_\mathrm{B}T\simeq 4.16V$ is relatively high compared with the characteristic energy scale of the system's dynamics, for which a large portion of quantum coherence in the energy eigenstates has been washed out in the canonical distribution. The remaining difference between magnitudes of $I_\mathrm{ca}$ and $I_\mathrm{ss}$ (especially for the strong system-environment coupling limit) should be attributed to the fact that the three independent environments, although with equal temperature, locally couple to the system at different sites in the network. It is the locality of the system-environment interaction that tends to destroy the quantum coherence between different sites that is inherent in the open system's energy eigenstates. As a result, the electronic current, which is proportional to the quantum coherence of electron at different sites (see the definition of the current operator below Eq.~\eqref{eq: current formula}), would be strongly suppressed by the decoherence, leading to the magnitude of $I_\mathrm{ss}$ smaller than that of $I_\mathrm{ca}$.

\end{appendix}

%%%%%%%%%%%%%%%%%%%%%%%%%%

%%%%%%%%%%%%%%%
\end{document}